\documentclass[10pt,twoside]{IEEEtran}

\renewcommand{\Pr}{\mathbb{P}}

\newcommand{\beq}{\begin{equation}}
\newcommand{\eeq}{\end{equation}}
\newcommand{\beqa}{\begin{eqnarray}}
\newcommand{\eeqa}{\end{eqnarray}}

\newcommand{\bA}{\mbox{\boldmath{$A$}}}

\newcommand{\VAR}{\textnormal{VAR}}

\newcommand{\sinc}{\textnormal{sinc}}
\newcommand{\mbbR}{{\mathbb{R}}}
\newcommand{\E}{\mathbb{E}}

\newcommand{\cS}{{\cal S}}
\newcommand{\cT}{{\cal T}}
\newcommand{\cU}{{\cal U}}
\newcommand{\cV}{{\cal V}}
\newcommand{\cW}{{\cal W}}
\newcommand{\cZ}{{\cal Z}}
\usepackage{amssymb}
\usepackage{mathrsfs}
\usepackage{graphicx}
\usepackage{epstopdf}

\usepackage[cmex10]{amsmath}

\begin{document}
%
\title{The Embedding Capacity of Information Flows Under Renewal Traffic}

\author{Stefano~Marano, Vincenzo~Matta, Ting~He, and Lang~Tong
\thanks{S.~Marano and V.~Matta are with DIIIE, University of Salerno,
via Ponte don Melillo I-84084, Fisciano (SA), Italy. E-mails: \{marano, vmatta\}@unisa.it.
T.~He is with IBM T.\ J.\ Watson Research Center, Hawthorne, NY. E-mail: the@us.ibm.com.
L.~Tong is with ECE Department, Cornell University, Ithaca, NY 14853 USA. E-mail: ltong@ece.cornell.edu.
}}

\maketitle

\begin{abstract}
Given two independent point processes and a certain rule for matching points between them, what is the fraction of matched points over infinitely long streams? In many application contexts, e.g., secure networking, a meaningful matching rule is that of a maximum causal delay, and the problem is related to embedding a flow of packets in cover traffic such that no traffic analysis can detect it. We study the best undetectable embedding policy and the corresponding maximum flow rate ---that we call the embedding capacity--- under the assumption that the cover traffic can be modeled as arbitrary renewal processes.
We find that computing the embedding capacity requires the inversion of very structured linear systems that, for a broad range of renewal models encountered in practice, admits a fully analytical expression in terms of the renewal function of the processes. Our main theoretical contribution is a simple closed form of such relationship. This result enables us to explore properties of the embedding capacity, obtaining closed-form solutions for selected distribution families and a suite of sufficient conditions on the capacity ordering. 
We evaluate our solution on real network traces, which shows a noticeable match for tight delay constraints. 
A gap between the predicted and the actual embedding capacities appears for looser constraints,  and
further investigation reveals that it is caused by inaccuracy of the renewal traffic model rather than of the solution itself. 
\looseness=-1
\end{abstract}

\section{Introduction}
\begin{figure}
\centerline{\includegraphics[height=.17\textheight]{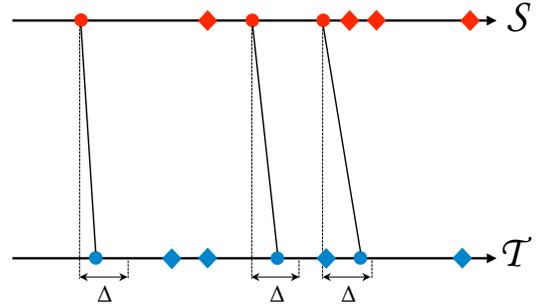}}
\caption{Notional sketch of the addressed problem, with arrival epochs of processes $\cS$ and $\cT$ matched according to a delay constraint $\Delta$. Matched points are marked by circles, unmatched by diamonds.}
\label{fig:open}
\end{figure}

\IEEEPARstart{C}{onsider} the pair of timing sequences represented by the point processes $\cS$ and $\cT$ in Fig.~\ref{fig:open}, where points are matched according to some prescribed rule.
What is the maximum achievable fraction of matched points (embedding capacity)  given the two processes and the matching rule?
How do statistical properties of the point processes affect the maximum fraction of matching?
The main theme of this paper is that of providing analytical tools for computing the embedding capacity of two independent and identically distributed renewal processes, when the coupling rule is formulated in terms of a causal delay constraint.

The above problem naturally arises in many applicative scenarios: from intelligence applications aimed at tracing relationships among individuals (e.g., in social networks), to the discovering of neuron connections by measurements of firing sequences, and so forth~\cite{Abbott-book,mainen-science}.
An application closer to the communication area concerns the anonymous relaying of messages in distributed architectures, or the detection of clandestine information flows in wireless systems. In fact, the evaluation of the embedding capacity under causal delay constraint has been recognized as a relevant problem in the context of secure networking, where the focus is on information flowing that is anonymous with respect to an attacking eavesdropper~\cite{Venk-He-Tong-IT}, or, in a reversed perspective, clandestine with respect to a legitimate traffic analyst~\cite{He-Tong-IT}.

In these contexts ---to which we specifically refer in the paper--- the two processes represent the sequences of time epochs (traffic patterns) at which successive packets leave two nodes of the network and, for security requirements, packets are encrypted so that they do not reveal special characteristics.
Still, the {\em act of transmission itself} cannot be kept secret, and timing analysis can be performed.

Given that nodes are unable to hide the act of transmission, they must hide the  information flow packets into their normal transmission scheduling, which provide {\em cover traffic} for the desired flow.
The  nodes can mask the timing relationships by properly delaying the transmission of information packets and/or multiplexing information packets with dummy packets or packets from other flows. 
With a sufficient amount of perturbation, an information flow can be disguised as traffic of arbitrary patterns.
In particular, the flow can appear identical to independent traffic following certain transmission schedules.

As a consequence, every transmission schedule (or cover traffic) has certain capacity of being utilized to transmit information flows covertly.
The matching capability of a particular schedule takes the operational meaning of an {\em embedding capacity}, that is, the maximum fraction of information packets that can be embedded in the cover traffic following this schedule, leaving no chances of discovering the presence of the flow itself.

	\subsection{Summary of Results}
The embedding capacity for a Poisson process under causal delay constraint is known, see~\cite{He-Tong-IT}. The Poisson assumption, however, rarely fits real traffic and, to date, analytical formulas for arbitrary renewal traffic are still missing. The contribution of this paper is in filling this gap.

We find that the embedding capacity is related to the invariant distribution of a certain Markov chain. 
First, we prove the existence of such distribution, so that capacity evaluation requires the solution of an integral equation.
We attack this problem by exploiting the powerful tools offered by the Riemann-Hilbert theory, which allows us
to derive the following approximation for the embedding capacity:
\[
C^\ast\approx
\frac{\lambda\Delta}{1+\displaystyle{\frac{2}{\lambda\Delta}\int_0^{\lambda\Delta} m(t)dt}}
\]
where $\lambda$ is the rate of the processes, $\Delta$ is the delay constraint, and $m(t)$ is the renewal function of the (scaled to unit rate) underlying process. 
The accuracy of this formula is excellent for a very broad range of renewal processes of interest for the applications, see Sect.~\ref{sec:appl}. 
We also show how $C^\ast$  can be computed to any degree of approximation by inverting a very structured linear system, and provide a  first-order correction expressed in closed form.

The significance of the above formula is that $C^\ast$ depends only on the renewal function which is the key quantity in renewal theory, as such, is well studied and understood. 
In many cases of practical interest, the integral involved can be evaluated explicitly, from which physical insights can be gained. 

The above expression is then used to relate the physical parameters and properties of the renewals to the embedding performance. 
In the asymptotic regime of large $\lambda\Delta$, the dispersion index $\gamma$ is the only relevant quantity, and the  capacity scales as $1-\gamma/(\lambda\Delta)$. 
Stochastic variability is instead the key (for any $\lambda\Delta$) to compare different interarrival distributions: less variable interarrivals yield a larger embedding capacity.

\subsection{Relevance to Secure Networking }
One applicative scenario of interest is that of secure networking.
Consider two packet streams in a network, whose transmission timestamps are represented by point processes $\cS=(S_1,S_2,\dots)$ and $\cT=(T_1,T_2,\dots)$. We assume that the packet content is fully protected by encryption, while the transmission patterns are relatively easy to obtain by a monitoring agent, which is capable of performing traffic (actually timing) analysis.

In one possible scenario, $\cS$ and $\cT$ are transmission activities of two nodes $N_1$ and $N_2$ in a wireless network. The existence of a flow from $\cS$ to $\cT$ implies that $N_2$ is acting as a relay for (part of the transmissions from) $N_1$, thus revealing a multi-hop route that is otherwise unobservable in protocol or content domain. The monitoring agent is interested in discovering whether or not such relaying exists; conversely, the nodes would like to hide the presence of the information flow (e.g., to preserve anonymity) by transmitting independently.

In another case, $\cS$ represents the pattern of the packets transmitted to a multiaccess relay through an ingoing link $L_1$, and $\cT$ is the pattern for an outgoing link $L_2$, both among multiple ingoing/outgoing links.
It is known that information is flowing from $L_1$ through the multiaccess relay, but it is not known whether or not the outgoing link $L_2$ is being used for this. The monitoring agent is interested in tracing the route of the flow, and the multiaccess relay intends to hide the route by scheduling the two links independently.

	\subsection{Related Work \& Organization}

The roots of packet embedding into cover traffic can be traced back to the early '80s.
The problem of avoiding traffic analysis using special relay policies was first considered in~\cite{Chaum}, with the adoption of the so-called MIX relays, that perform multiplexing, scrambling and encryption of the incoming traffic in order to eliminate the correlation with the outgoing traffic.
Since then, several studies have been made in order to improve relay performances, see e.g.~\cite{Hajek,ZhuGraham}.
More recently, it has been shown how statistically independent transmission schedules can achieve perfectly anonymous relaying, with emphasis on the maximization of the carried information capacity~\cite{Venk-He-Tong-IT}.

Also related to our problem is the network security issue  referred to as stepping-stone attack~\cite{donohoetal-intrusion,StanifordHeberlein}, in which an adversary launches an attack through a sequence of compromised servers, and one would like to trace the sequence to the origin of the attack.  For wireless networks, an ad hoc network may be subject to the worm-hole attack~\cite{wormhole}, where the attacker hijacks the packets of a node and channels them through a covert tunnel. In such scenarios, the maximum information rate sustainable by the attackers is related to the embedding capacity of the node traffic patterns.

From an information theoretic perspective, the problem of secure communications, in terms of maximizing the reliable rate to a legitimate receiver with secrecy constraints with respect to an eavesdropper, has been extensively studied, since the pioneering works~\cite{shannon-secrecy,Wyner,CK-broadcast}, up to recent extensions, including multiaccess~\cite{Liang-Poor-IT}, fading~\cite{Liang-Poor-Shamai-IT}, feedback~\cite{FranceschettiWiretap}, and broadcast ~\cite{BiaoChenBroadcast} channels, among many others.
We stress that the specific scenario of interest for this paper is instead  secure networking with focus on anonymous relaying of information, according to the model proposed in~\cite{Venk-He-Tong-IT,He-Tong-IT}.

Formal studies of the embedding properties of renewals have been carried out in~\cite{Venk-He-Tong-IT,He-Tong-IT} , with extensions to distributed detection with  communication constraints~\cite{He-Tong-Forens,He-Agaskar-Tong-SP}.
The authors of~\cite{He-Tong-IT} settled up the problem from the traffic analyzer's perspective, where the role of the embedding capacity is replaced by that of undetectable flow. They found a closed formula for the capacity under the Poisson regime.
General renewal traffic models in many applications (inside the communication area as well as outside that) are far from being approximated as Poisson, such that several extensions of the above studies in this direction have been proposed, see~\cite{HeTongICASSP2008,HeTongSwamiALLERTON2009}. 
However, a tractable analytical formula for the embedding capacity under arbitrary renewal traffic is still missing.

The remainder of this paper is organized as follows. 
Section~\ref{sec:form} formalizes the problem, the main results of the paper are presented in Sect.~\ref{sec:main}, and 
Sect.~\ref{sec:proof} is devoted to the main mathematical derivations.
Sect.~\ref{sec:appl} concerns the application of the main theoretical findings to specific examples, while Sect.~\ref{sec:ordering} addresses the problem of classification and ordering of renewal processes in terms of their embedding capacity.
Finally, Sect.~\ref{sec:realtraces} presents the results of experiments on real network traces, and conclusions follow in Sect.~\ref{sec:concl}. An appendix contains some mathematical derivations.

\section{Problem Formulation}
\label{sec:form}

Capital letters denote random variables, and the corresponding lowercase the associate realizations, while $\Pr$ and $\E$ denote probability and expectation operators, respectively.

Consider two point processes $\cS=(S_1,S_2,\dots)$ and $\cT=(T_1,T_2,\dots)$ defined over the semi-axis $t \in (0, \infty)$. Points that are matched over the two processes form an {\em information flow} in the sense that one point in a matched pair can be thought of as a relayed copy of the other. We are interested in delay-sensitive directional flows, for which matched points obey a causal bounded delay constraint as follows.

\vspace*{5pt} \noindent
\textsc{Definition 1}
{\em (Information flow) Processes $\cW=(W_1,W_2,\dots)$ and $\cZ=(Z_1,Z_2,\dots)$ form a $\Delta$-bounded-delay information flow in the direction $\cW \rightarrow \cZ$ if for every realization $\{w_i\}$ and $\{z_i\}$, there is a one-one mapping
$\{w_i\}\rightarrow\{z_i\}$ that satisfies the causal bounded delay constraint
$0\leq z_i-w_i\leq\Delta$, $\forall i$.}~\hfill$\diamond$\looseness=-1

\vspace*{2pt} \noindent
Here $\Delta>0$ is a known constant representing the maximum tolerable delay during relaying.

Given point processes $\cS=(S_1,S_2,\dots)$ and $\cT=(T_1,T_2,\dots)$, an information flow can be generated by finding, for each realization of the processes, subsequences that admit a valid one-one mapping. This is controlled by an embedding policy.

\vspace*{5pt} \noindent
\textsc{Definition 2}
{\em (Embedding policy) An embedding policy $\epsilon$ selects subsequences $\cW^\epsilon$ of $\cS$ and $\cZ^\epsilon$ of $\cT$ to form an information flow.}~\hfill$\diamond$

\vspace*{2pt} \noindent
The name ``embedding'' is due to the fact that to an outsider who cannot observe the selection, it is  not known which points belong to an information flow or even if there is a flow, and thus the flow is embedded in the overall processes $(\cS,\: \cT)$. For the same reason, $(\cS,\: \cT)$ is called \emph{cover traffic}.

Let ${\cal E}=\{\epsilon\}$ be the set of admissible embedding policies. Given $\epsilon \in {\cal E}$, the cover traffic $(\cS,\: \cT)$ is decomposed into
\[
\cS=\cW^\epsilon\oplus \cU^\epsilon,\quad
\cT=\cZ^\epsilon\oplus \cV^\epsilon,
\]
where $(\cW^\epsilon,\cZ^\epsilon)$ forms a valid information flow. Here $\oplus$ is the superposition operator for point processes: $\{ c_i\}$=
$\{ a_i\} \oplus \{ b_i\}$ means that $\{ c_i\}=\{ a_i\} \cup \{ b_i\}$ with $c_1 \le c_2 \le \dots$.

Given the cover traffic, each embedding policy has a certain capability of hosting information flows, quantified as follows.

\vspace*{5pt} \noindent
\textsc{Definition 3}
{\em (Efficiency)
Given cover traffic $(\cS,\: \cT)$, the efficiency of an embedding policy $\epsilon\in {\cal E}$ is measured by
\[
\eta(\epsilon):=\lim_{t\rightarrow\infty}\frac{\,N_{\cW^\epsilon}(t)+N_{\cZ^\epsilon}(t)}{N_{\cS}(t)+N_{\cT}(t)},
\]
where $N_{\cW^\epsilon}(t)$, $N_{\cZ^\epsilon}(t)$ are the counting processes for the embedded information flow, so are $N_{\cS}(t)$, $N_{\cT}(t)$ for the cover traffic (assuming the limit exists almost surely).}~\hfill$\diamond$

\vspace*{2pt} \noindent
That is, the efficiency is the asymptotic fraction of matched points in the cover traffic. We are interested in the highest efficiency that we call the {\em embedding capacity}.

\vspace*{5pt} \noindent
\textsc{Definition 4}
{\em (Embedding capacity)
$
C^\ast=\sup_{\epsilon\in{\cal E}} \eta(\epsilon).
$}~\hfill$\diamond$

\vspace*{5pt} \noindent
The embedding capacity $C^*$ is a function of the cover traffic and the flow constraints (e.g., $\Delta$), omitted for simplicity.
We shall focus on the case that the cover traffic $\cS$ and $\cT$ are independent and identically distributed (i.i.d.) renewal processes, with interarrival random variables $X$ and $Y$, respectively.
Throughout the paper it is assumed  that $X$ and $Y$ are absolutely continuous with known Probability Density Function (PDF) $f(t)$ and Cumulative Distribution Function (CDF) $F(t)$, and that the rate of the processes, denoted by $\lambda$, is finite and nonzero, i.e., $0<\lambda = 1/\E[X]=1/\E[Y]< \infty$.
When the second moment is finite, we define the dispersion index as
\beq
\gamma=\lambda^2\,\VAR[X]=\lambda^2\,\VAR[Y]<\infty.
\label{eq:dispindx}
\eeq

\begin{figure*}
\centerline{\includegraphics[height=.3\textheight]{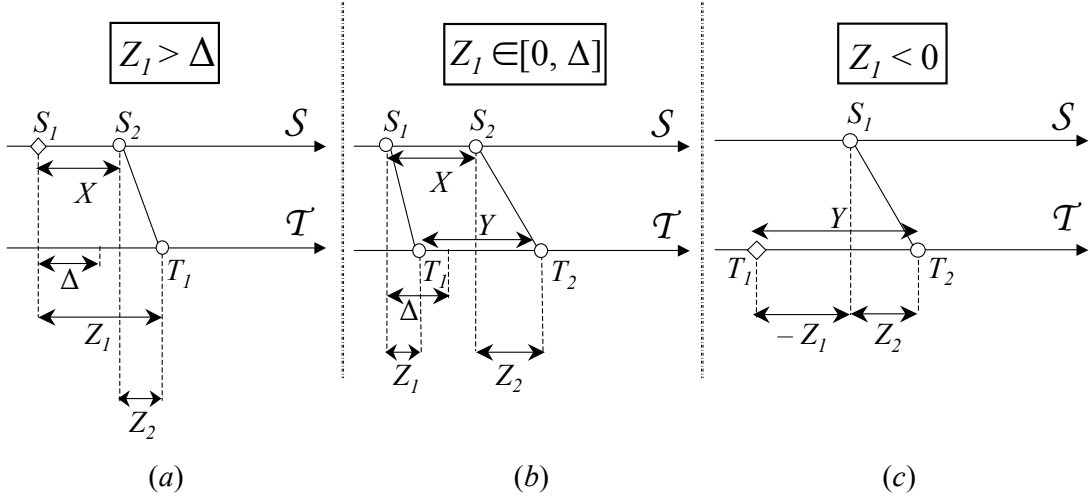}}
\caption{
Three situations arising from applying the BGM procedure to point processes ${\cal S}$ and ${\cal T}$. Chaff points are denoted by ``$\diamond$''. \textit{Left:} the point at $S_1$ is unmatched, and it is a chaff point in the first process. \textit{Center:} all points are matched (no chaff). \textit{Right:} a chaff point is present in the second process.}
\label{fig:scheme}
\end{figure*}

\section{Characterization of the Embedding capacity}
\label{sec:main}

	\subsection{Optimal Embedding Policy}

As a first step toward embedding capacity evaluation, we need to find an optimal embedding policy that maximizes the number of matched points for any given cover traffic, thus achieving the embedding capacity. This has been achieved by an existing algorithm called the \emph{Bounded Greedy Match (BGM)}~\cite{Blumetal}.
It is a simple algorithm that classifies the points of two arbitrary point processes as ``matched'' and ``unmatched'' by sequentially matching points in the two processes under a causal delay constraint $\Delta$ and marking the points violating this constraint as unmatched.

The BGM algorithm works as follows. Given realizations of two point processes, all the points initially ``undetermined'', the BGM repeats the following steps, see Fig.~\ref{fig:open}: 
\begin{enumerate}
	\item Consider the first (in the direction of increasing time) undetermined point in the first process, say $p^{(1)}$;
	\item Find the first undetermined point in the second process in the interval $[p^{(1)},\: p^{(1)}+\Delta]$, if any, denoted by $p^{(2)}$;
\item If such a point exists, mark both $p^{(1)}$ and $p^{(2)}$ as ``matched''; otherwise, mark $p^{(1)}$ as ``unmatched''; in either case, mark all undetermined points in the second process before $p^{(1)}$ ``unmatched''.
\end{enumerate}
Matched and unmatched points are also referred to as ``flow'' and ``chaff'', respectively.
The BGM is optimal in the sense that, given two arbitrary realizations of point processes and an arbitrary value of $\Delta$, 
the algorithm finds the maximum number of matched points satisfying the delay bound~\cite{Venk-He-Tong-IT,He-Tong-IT,Blumetal}.

	\subsection{Embedding capacity in terms of a Markov Chain}

Our second step in deriving the embedding capacity $C^\ast$ consists of modeling the behavior of BGM by a Markov chain, whose stationary distribution is directly related to $C^\ast$.

With reference to Fig.~\ref{fig:scheme}, let us consider the time difference between the first points, that is
$Z_1=T_1-S_1$.
According to the BGM algorithm, we have the following three possibilities:

\vspace*{2pt} \noindent $(i)$ If $Z_1>\Delta$, the points cannot be matched, and the one in $\cS$ is labeled as chaff. To decide the nature (chaff/non chaff) of the point in $\cT$, we must check whether it can be matched to the next arrival in $\cS$, thus computing, see Fig.~\ref{fig:scheme}$(a)$,
\[
Z_2=T_1-S_2=Z_1- X,
\]
where $X$ is the random variable representing the interarrivals in $\cS$.

\vspace*{2pt} \noindent $(ii)$ If $0\leq Z_1\leq\Delta$, the points match. To check the nature of the next incoming points, we update the process as, see Fig.~\ref{fig:scheme}$(b)$,
\[
Z_2=T_2-S_2=Z_1+Y- X,
\]
where $Y$ is the random variable representing the interarrivals in $\cT$ (recall that $Y$ has the same distribution as $X$).

\vspace*{2pt} \noindent $(iii)$ If $Z_1<0$, the points cannot be matched, and the one in $\cT$ is labeled as chaff. To decide the nature of point in $\cS$, we must check whether it can be matched to the next arrival in $\cT$, thus computing, see Fig.~\ref{fig:scheme}$(c)$,
\[
Z_2=T_2-S_1=Z_1+Y.
\]

By repeating for the successive points, we see that a Markov process can be compactly defined in terms of the original renewals by the following recursion rule
\beq
Z_n=
\left\{
\begin{array}{lll}
Z_{n-1} - X_n,\qquad &\mbox{if }\; Z_{n-1}>\Delta,\\
Z_{n-1} + Y_n- X_n,\qquad &\mbox{if }\; 0\leq Z_{n-1}\leq\Delta,\\
Z_{n-1} +  Y_n,\qquad &\mbox{if }\; Z_{n-1}<0,
\end{array}
\right.
\label{eq:Markov}
\eeq
where $X_n$ and $Y_n$ are the interarrivals of the first and the second process at the $n$th step of the chain,  following the common PDF $f(t)$.

\begin{figure}
\centerline{\includegraphics[height=.3\textheight]{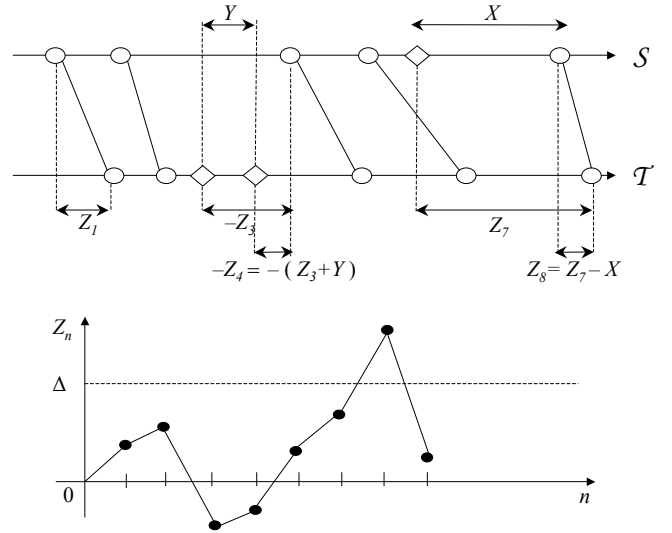}}
\caption{Construction of a sample path of the Markov process (lower panel) from a realization of the two point processes (upper). In the upper panel, the points marked with ``$\diamond$'' are those classified as chaff by the BGM algorithm.}
\label{fig:walk}
\end{figure}

The Markov chain defined in~(\ref{eq:Markov})
is schematically illustrated in Fig.~\ref{fig:walk}.  According to the constitutive equation (\ref{eq:Markov}), the increment of the Markov chain is the interarrival difference $Y - X $ if the chain is currently between $0$ and $\Delta$, which implies the matching is successful (e.g., $Z_1$, $Z_8$); the increment is $Y$ if the chain is below $0$, in which case the reference point in the second process is marked as chaff and the reference point in the first process remains the same (e.g., $Z_3$, $Z_4$); similarly, the increment is $-X$ if the chain is above $\Delta$, when the reference point in the first process becomes chaff and that in the second process remains unchanged (e.g., $Z_7$).
Note that the number of steps of the Markov chain lying inside (resp. outside) the barriers $0$ and $\Delta$ defines the number of flow (resp. chaff) points marked by the BGM algorithm. This suggests that a relationship exists between the asymptotic distribution of the chain and the fraction of flow points, i.e., the embedding capacity. This is made precise in the next section.

	\subsection{Main Results}

The first theorem we present, whose proof is deferred to appendix~\ref{app:Markov}, establishes a connection between the  embedding capacity and the  invariant distribution of the BGM Markov chain, expressed as the solution of an integral equation.

\vspace*{5pt} \noindent
\emph{\textsc{Theorem 1} ($C^\ast$ by Markov chain) Let $\cS$ and $\cT$ be two independent and identically distributed renewal processes, with interarrival PDF $f(t)$. Let $\Delta$ be the delay constraint, and define a Markov chain by~(\ref{eq:Markov}). Assume BGM can match at least one pair of points in $\cS$ and $\cT$ almost surely.}

\vspace*{2pt}\noindent $a)$ \emph{The invariant PDF $h(t)$ of the Markov chain exists and solves the following homogeneous Fredholm integral equation of the second kind~\cite{Pipkin}
\beqa
h(t)&=&\int_{-\infty}^{0} h(\tau) f(t-\tau) d\tau+\int_{\Delta}^{+\infty} h(\tau) f(\tau-t) d\tau\nonumber\\
&+&\int_{0}^{\Delta} h(\tau) f_0(t-\tau) d\tau,
\label{eq:mainEq0}
\eeqa
where $f_0(t)$ is the convolution between $f(t)$ and $f(-t)$, defined as $\int_0^{+\infty} f(\tau) f(\tau-t)d\tau$.}

\vspace*{2pt}\noindent $b)$ \emph{The embedding capacity can be written as
\beq
C^\ast=\frac{2\int_{0}^{\Delta}h(t)dt
}{1+\int_{0}^{\Delta}h(t)dt
}.
\label{eq:CapExpress}
\eeq}~\hfill$\diamondsuit$

Since now, we shall assume that the hypotheses of Theorem~1 are in force. The next theorems, whose proofs are given in Sect.~\ref{sec:proof}, accordingly focus on the solution of the integral equation relevant to capacity computation. 
We first introduce the following definitions.

\vspace*{5pt} \noindent
\textsc{Definition 5}
{\em (u-PDF) The probability density function $k(t)$ of the interarrivals scaled to unit mean, that is, the random variables $\lambda X$ and $\lambda Y$, will be called u-PDF.
}~\hfill$\diamond$

\vspace*{5pt} \noindent
\textsc{Definition 6}
{\em (u-RF) The Renewal Function
\[
m(t):=\E[N(t)],
\]
where $N(t)$ is the number of arrivals in $(0,t)$ of the  processes scaled to unit rate, having interarrival random variables $\lambda X$ and $\lambda Y$, will be called u-RF.
}~\hfill$\diamond$

\vspace*{5pt} \noindent
\emph{\textsc{Theorem 2} (Exact value of $C^\ast$)
Under the assumption of finite second moment for the interarrivals, the embedding capacity of two independent and identically distributed renewal processes with rate $\lambda$, under delay constraint $\Delta$, is
\beq
C^\ast=\displaystyle{\frac{2\Omega(0)}{1+\Omega(0)}},
\label{eq:inthethe}
\eeq
where $\Omega(f)$ is the solution of
\beqa
\Omega(f)+
2\,\int{\Omega(\nu)\Re\left\{\frac{K(\nu)}{1-K(\nu)} \right \}
\lambda\Delta\sinc[\lambda\Delta(f-\nu)]d\nu}
\nonumber \\
=\lambda\Delta\sinc(\lambda\Delta\,f)\,\frac{1-\Omega(0)}{2},
\label{eq:finalfinal0}
\eeqa
$K(f)$ being the Fourier transform of the u-PDF~$k(t)$, and $\sinc(t)=\sin(\pi\,t)/(\pi t)$}.~\hfill$\diamondsuit$

\vspace*{5pt} \noindent

As a check, let us specialize the above equation to the case of exponential interarrivals, for which embedding capacity is available in closed form~\cite{He-Tong-IT}. It is easily seen that  $\Re\left\{\frac{K(f)}{1-K(f)}\right\}=0$, allowing direct solution of eq.~(\ref{eq:finalfinal0}), and computation of $\Omega(0)=\lambda\Delta/(2+\lambda\Delta)$. Substituting into eq.~(\ref{eq:inthethe}), this yields
\[
C^\ast=\frac{\lambda\Delta}{1+\lambda\Delta} \qquad \textnormal{(exponential)},
\]
that matches the known result from~\cite{He-Tong-IT}.

Note that Theorem~2 still gives an implict solution to the problem in terms of an integral equation,  which does not have a closed-form solution in general.
On the other hand, eq.~(\ref{eq:finalfinal0}) turns out to be amenable to approximate solutions, thus yielding the results stated in the next two theorems.

\vspace*{5pt} \noindent
\emph{\textsc{Theorem 3} (Approximation of $C^\ast$) Under the assumption of finite second moment for the interarrivals, the embedding capacity of two independent and identically distributed renewal processes with rate $\lambda$, under delay constraint $\Delta$, can be approximated as
\beqa
C^\ast\approx C= \displaystyle{\frac{\lambda\Delta}{1+\displaystyle{\frac{2}{\lambda\Delta}\int_0^{\lambda\Delta} m(t)dt}}},
\label{eq:finaltheorem}
\eeqa
$m(t)$ being the u-RF.~\hfill$\diamondsuit$}

\vspace*{5pt} \noindent

Again, let us apply eq.~(\ref{eq:finaltheorem}) in the Poisson regime. The u-RF of an exponential random variable  is $m(t)= t$, that inserted in~(\ref{eq:finaltheorem}) gives
\[
C=\frac{\lambda\Delta}{1+\lambda\Delta} \qquad \textnormal{(exponential)},
\]
implying that, in this particular case, formula~(\ref{eq:finaltheorem}) is exact, i.e., $C^\ast=C$.
This can be understood by looking at the technique used to get the approximation\footnote{The terms $k \neq 0$ neglected in eq.~(\ref{eq:mequalzero}), are rigorously zero in the exponential case, since the integral in~(\ref{eq:symm0}) is zero.} in the proof of Theorem~3.

The relevance of Theorem~3 stems from the fact that, for the typical interarrival distributions encountered in many applications, the accuracy of the fully analytical approximation~(\ref{eq:finaltheorem}) seems to be excellent irrespective of the range of the product $\lambda\Delta$, the tailweight of the distribution, its variance (and even for infinite second moment), as confirmed by the examples in Sect.~\ref{sec:appl}.
Accordingly, Theorem~3 provides us with an accurate and mathematically tractable expression for the embedding capacity under arbitrary renewal traffic.

We would like to emphasize that the characterization~(\ref{eq:finaltheorem}) relates the sought capacity to the u-RF of the underlying process. This highlights the role of the renewal function $m(t)$, and reveals
that its average $\frac 1 {\lambda\Delta}\int_0^{\lambda\Delta} m(t) dt$ is the key quantity in determining $C$. Thus, different traffic models can be classified with respect to their embedding capabilities just in terms of that average.

We now state a corollary characterizing the asymptotic behavior of the capacity in the limit of  $\Delta\gg 1/\lambda$.
From a known property of the renewal function~\cite{ross-book}, $m(t)- t \rightarrow (\gamma-1)/2$ in the limit of $t \rightarrow \infty$, where $\gamma$ is the dispersion index defined in eq.~(\ref{eq:dispindx}). Simply plugging that expression in eq.~(\ref{eq:finaltheorem}) would give $1-C\sim \gamma/(\lambda\Delta)$. Indeed, we have the following result.

\vspace*{5pt} \noindent
\emph{\textsc{Corollary 1} (Scaling law for $C$) Under the assumption of finite second moment for the interarrivals,
$\lim_{\lambda\Delta \rightarrow \infty} [1-C] (\lambda\Delta) =\gamma$, i.e., the embedding capacity in Theorem 3 scales as
\[
1-C\sim \frac{\gamma}{\lambda\Delta}.
\]}~\hfill$\diamondsuit$

The corollary reveals that, for large values of the product $\lambda\Delta$, the key quantity in determining the capacity is the dispersion index: given $\lambda\Delta\gg 1$, the ability for a type of (renewal) traffic to hide information flows in independent realizations only depends on the value of the dispersion index $\gamma$, and different traffic models sharing the same dispersion index behave similarly.

Finally, to improve on the approximation in Theorem~3, we provide the following theorem that expresses the embedding capacity as the solution to a simple linear system.
Consider, for any integer $N \ge 1$, the following system
\[
\sum_{k=-N}^N A_{hk} \, \Omega\left(\frac{k}{\lambda\Delta}\right)=\frac{\lambda\Delta} 2 I_h, \qquad h=-N,\dots,N,
\]
where $I_h=1$ for $h=0$, and $I_h=0$ otherwise.
The analytical expressions of the entries $A_{hk}$, defining a $2N+1$ by $2N+1$ matrix $\bA$, are
\beqa
&&\hspace*{-40pt}A_{00}=
1-\frac {\lambda\Delta}{2} +\displaystyle{\frac{2}{\lambda\Delta}\int_0^{\lambda\Delta} m(t)dt },
\label{eq:A00} \\
&&\hspace*{-40pt}A_{kk}=
1+ \frac {2} {\lambda\Delta} \int_0^{\lambda\Delta} m(t) \left[\cos\left(\frac{2\pi k t}{\lambda\Delta}\right)\right.\nonumber\\
&&\left.+\,  2 \pi k \, \left( 1-\frac{t}{\lambda\Delta}\right)\, \sin \left(\frac{2\pi k t}{\lambda\Delta}\right) \right] \, dt ,
~~ k\neq 0, \label{eq:Akk} \\
&&\hspace*{-40pt}A_{0k}=
\frac{2\,(-1)^{k}}{\lambda\Delta}\int_0^{\lambda\Delta} m(t) \,\cos\left(\frac{2\pi k t}{\lambda\Delta}\right) \, dt,
~~~~~~~~ k \neq 0, \label{eq:symm0} \\
&&\hspace*{-40pt}A_{hk}=
\frac{(-1)^{h-k}}{(h-k)} \left [ h (-1)^h A_{0h} -  k (-1)^k A_{0k}\right ],
~~~h \neq k. \label{eq:symm}
\eeqa

\vspace*{5pt} \noindent
\emph{\textsc{Theorem 4} (Linear system approximation of $C^\ast$) Under the assumption of finite second moment for the interarrivals, let
$C^\ast=\frac{2\Omega(0)}{1+\Omega(0)}$ as in Theorem 2. Then, assuming that $\bA$ is invertible, $\Omega(0)$ can be approximated as $\lambda\Delta/2$ times the $(0,0)$-entry of matrix $\bA^{-1}$, namely
$\Omega(0)=\frac{\lambda\Delta}{2} \, \{\bA^{-1}\}_{00}$.
In particular, specializing for $N=1$, the capacity becomes
\beq
C^\ast\approx \displaystyle{
\frac{\lambda\Delta}{
1+\displaystyle{
\frac 2 {\lambda\Delta}\,\int_0^{\lambda\Delta} m(t)dt + 2\,\frac{A_{01}^2}{A_{01}-A_{11}}}} \, .
}
\label{eq:lineartheorem}
\eeq
}~\hfill$\diamondsuit$

\noindent
\emph{\textsc{Remark 1}.}
Note that, in the approximation corresponding to $N=1$, a correction term $2\,\frac{A_{01}^2}{A_{01}-A_{11}}$ appears, with respect to  $C$ in eq.~(\ref{eq:finaltheorem}), which  uses only $A_{00}$.
Also, from eqs.~(\ref{eq:A00})--(\ref{eq:symm}) we see that $\bA$ is very structured and its degrees of freedom grow only linearly with $N$; in fact, $\bA$ is completely specified by assigning one row and the main diagonal. This structure is very convenient for numerical tractability.
Finally, it is expected that the solution becomes more and more accurate as the system size $N$ increases. In the section devoted to numerical experiments, we show that the {\em zero-order} approximation $C$ is well satisfying in many cases of interest. Even when this is not strictly true, a first-order correction~(\ref{eq:lineartheorem}) offers very good results.

\noindent
\emph{\textsc{Remark 2}.}
Let us consider a random variable with u-PDF $k(t)$ which is zero for $t<a$, some $a>0$. We have  $m(t)=0$ for $t<a$.
This implies that, in the range $\lambda\Delta<a$, the cross terms $A_{0k}$ with $k\neq 0$ vanish, so that the approximation~(\ref{eq:finaltheorem}) is exact, and gives the linear relationship $C^\ast=\lambda\Delta$ in the considered range, as verified later\footnote{The same conclusion can be also argued as follows. For delay $\Delta$ smaller than the minimum allowed interarrival time, $S_k-S_{k-1}>\Delta$, such that the probability that $S_k$ matches is  the probability that the first arrival after $S_k$ in $\cT$ occurs before $S_k+\Delta$ and it can be computed, due to independence between the processes, by using the residual lifetime distribution~\cite{ross-book}:
$\Pr[S_k \textnormal{ matches}]\approx\lambda\int_0^\Delta [1-F(t)]dt$. This implies $\Pr[S_k \textnormal{ matches}]\approx \lambda\Delta$, for $\lambda\Delta<a$. By ergodicity, $C^\ast=\lambda\Delta$ in the considered range.} (see Fig.~\ref{fig:ShiftedExpo}). This is also consistent with earlier approximation and simulation results in~\cite{HeTongSwamiALLERTON2009}.

\section{Proofs of Theorems 2-4 via Riemann-Hilbert Theory}
\label{sec:proof}

In the following, we make use of suitable normalization of the relevant physical quantities, aimed at simplifying the mathematical derivation.
Indeed, the problem possesses a natural {\em scale-free} property. For a given distribution of the interarrival process, we note that doubling the arrival rate ``speeds up'' the system so that the sample paths can be redrawn on a time axis scaled by a factor $2$, and halving $\Delta$ leaves unchanged the number of matches.
We accordingly introduce the {\em normalized delay} $\delta=\lambda\Delta$, and work in terms of the unit-mean random variables with u-PDF $k(t)$.
It is further convenient to  symmetrize the problem by shifting the Markov chain, as well as the corresponding boundaries, which yields to the following Fredholm equation
\beqa
u(t)&=&\int_{-\infty}^{-\delta/2} u(\tau) k(t-\tau) d\tau+\int_{\delta/2}^{+\infty} u(\tau) k(\tau-t) d\tau\nonumber\\
&+&\int_{-\delta/2}^{\delta/2} u(\tau) k_0(t-\tau) d\tau,
\label{eq:mainEq}
\eeqa
where $k_0(t)$ is the convolution between $k(t)$ and $k(-t)$.

Equation~(\ref{eq:mainEq}) is a homogeneous Fredholm equation of the second kind, and we have three different regions where the integrals look like convolutions. Were the integral equivalent to a convolution as a whole, a simple transform  method would be directly applicable.
To elaborate, let us first consider what would happen if the function was known within the strip $[-\delta/2,\delta/2]$. In this case, the equation would be nonhomogeneous, and will be classified as a convolution-type equation with two distinct kernels. For this case a powerful approach, that can be traced back to Carleman and to Wiener and Hopf, still prescribes transforming the equation in the Fourier domain~\cite{PolyaninManzhirov} .
After transformation, the problem falls in the class of the so-called Riemann-Hilbert boundary value problems.

The Riemann-Hilbert problem,\footnote{Actually, there has been some uncertainty about the original pioneer of the approach. According to Muskhelishvili~\cite{Muskhelishvili} ``\emph{The problem formulated above is often called the Riemann problem, but the Author considers	 this name to be incorrect [\dots], because it was first considered by D.\ Hilbert essentially in the form in which it is stated}.''} in a nutshell, consists in finding two functions, analytic in the upper and lower half planes, respectively, whose difference on the real axis equals a known function~\cite{PolyaninManzhirov}~\cite{Muskhelishvili}. Direct application of this approach would require
that the sought stationary distribution be known within $[-\delta/2,\delta/2]$, but this is not our case. 	
Generalizations of the method have been proposed. They include the Carleman-Vekua regularization method, which suggests to initially treat the unknown function as known,
and formulating a new integral equation in terms of the function in the interval $[-\delta/2,\delta/2]$, and the work by Jones~\cite{Jones}, see also~\cite{Noble}. The proof that follows is based on these approaches.

Before, we need some basic notation and concepts about one-sided functions and their analytic Fourier transforms, which will be useful in the following.
For a generic function $g(t)$, let $g^{+}(t)=g(t)\,1(t)$ and $g^{-}(t)=-g(t)\,1(-t)$, where $1(t)$ is the Heaviside unit-step function $1(t)=1$ for $t>0$, and $1(t)=0$ for $t<0$.
In the Fourier domain, this means $G^+(f)=\int_{0}^{+\infty} g(t)e^{i2\pi f t}dt$, and
$G^-(f)=-\int_{-\infty}^0 g(t)e^{i2\pi f t}dt$.

By replacing the real parameter $f$ by a complex variable $z = f + iy$, the above integrals become
$G^{+}(z)=\int_{0}^{+\infty} g(t)e^{i2\pi z t}dt$ and $G^{-}(z)=-\int_{-\infty}^{0} g(t)e^{i2\pi z t}dt$,
which are analytic in those regions of the complex plane of the variable $z$ in which they are absolutely convergent~\cite{PolyaninManzhirov}: $G^{+}(z)$ is analytic for $\Im(z)>0$, and $G^{-}(z)$ for $\Im(z)<0$.

From Sokhotski-Plemelj formula~\cite{PolyaninManzhirov}, or simply decomposing the Fourier integral into the left and the right part, we have that, on the real axis,
\beq
G^{+}(f)=\frac{G(f) + i\, {\cal H}[G(f)]}{2}, \,
G^{-}(f)=\frac{-G(f) + i\, {\cal H}[G(f)]}{2},
\label{eq:ULHPsplit}
\eeq
where the Hilbert transform ${\cal H}[G(f)]=\frac{1}{\pi}\int \frac{G(\nu)}{f-\nu}d\nu$ has been introduced
(the integral is in the sense of Cauchy principal value).

\subsection{Proof of Theorem 2}
Consider the the unknown function $u(t)$ in eq.~(\ref{eq:mainEq}) and let us define
\[
u(t)=v^{+}(t-\delta/2) -v^{-}(t+\delta/2) + \omega(t),
\]
where
\beqa
v^{+}(t-\delta/2)&=&u(t)\,1(t-\delta/2),\nonumber\\
v^{-}(t+\delta/2)&=&-u(t)\,1(-t-\delta/2),\nonumber\\
\omega(t)&=& \left \{ \begin{array}{ll} u(t) & \delta/2 \le t \le \delta/2 \\ 0 & \textnormal{otherwise}\end{array} \right . .
\nonumber
\eeqa
The corresponding Fourier transforms will be accordingly denoted by $V^{+}(f)$, $V^{-}(f)$ and  $\Omega(f)$.
Note that, from~(\ref{eq:CapExpress}), we are just interested in $\int_{0}^{\Delta} h(t) dt=\int_{-\delta/2}^{\delta/2} u(t) dt=\Omega(0)$.

Transforming both sides of the integral equation~(\ref{eq:mainEq}) into the Fourier domain gives
\beqa
\lefteqn{V^{+}(f)e^{i\pi\delta f}-V^{-}(f)e^{-i\pi\delta f}+\Omega(f)}\nonumber\\
&=&V^{+}(f)e^{i\pi\delta f} \bar{K}(f)
-V^{-}(f)e^{-i\pi\delta f} K(f)\nonumber\\
&+& \, \Omega(f)|K(f)|^2,\nonumber
\eeqa
where $\bar{a}$ is the conjugate of $a$. The above equation can be recast as
\beq
\frac{V^{+}(f)e^{i\pi\delta f}}{1-K(f)}=\frac{V^{-}(f)e^{-i\pi\delta f}}{1-\bar{K}(f)} -W(f),
\label{eq:primaeq}
\eeq
where we define
\beqa
W(f)=\Omega(f)\frac{1-|K(f)|^2}{|1-K(f)|^2} =
\Omega(f)\left[1+2\,\Re\left\{\frac{K(f)}{1-K(f)} \right \}\right].
\nonumber\\
\label{eq:Xdef}
\eeqa
Multiplied by $e^{-i\pi\delta f}$, eq.~(\ref{eq:primaeq}) becomes
\[
\frac{V^{+}(f)}{1-K(f)}=\frac{V^{-}(f)e^{-i 2\pi\delta f}}{1-\bar{K}(f)} - W(f)e^{-i \pi\delta f},
\]
and using the factorization in~(\ref{eq:ULHPsplit}):
\[
W(f) e^{-i \pi\delta f}=[W(f) e^{-i \pi\delta f}]^{+}-[W(f) e^{-i \pi\delta f}]^{-}.
\]
Combining the above equations gives
\beqa
\lefteqn{\frac{V^{+}(f)}{1-K(f)}
+
{[W(f) e^{-i \pi\delta f}]}^{+}}
\nonumber\\
&&=
\frac{V^{-}(f)e^{-i 2\pi\delta f}}{1-\bar{K}(f)}
+
{[W(f) e^{-i \pi\delta f}]}^{-}.
\label{eq:RH}
\eeqa
By construction
the function $\frac{V^{+}(z)}{1-K(z)}$
is analytic in the upper half plane $\Im\{z\}>0$, continuous on the real axis, with a single pole located at $z=0$.
By the known property of the characteristic function, $K^\prime(0)=i 2\pi$, such that this pole has order one.
On the other hand, the function $\frac{V^{-}(f)e^{-i 2\pi\delta f}}{1-\bar{K}(f)}$ is analytic in $\Im\{z\}<0$, continuous on the real axis, with a single pole of order one located at $z=0$.
Similar considerations apply to $[W(f)e^{-i\pi\delta f}]^+$ and $[W(f)e^{-i\pi\delta f}]^-$, with the further property that there are no poles.\footnote{Note that, under the assumption of finite second moment,
\[
\lim_{f\rightarrow 0}\Re\left\{\frac{K(f)}{1-K(f)}\right\}=\frac{\gamma-1}{2},
\]
where the last equality straightforwardly follows by repeated application of De L'H$\hat{\textnormal{o}}$pital rule, and from $K^{\prime}(0)=i 2\pi$ and $K^{\prime\prime}(0)=-4\pi^2 (1+\gamma)$.
This ensures that $W(f)$ is well-behaved.
}

The asymptotic behavior of the involved functions is essentially determined by Fourier transforms, such that we assume boundedness at infinity.

Summarizing, the LHS and RHS  of eq.~(\ref{eq:RH}) define functions that are analytic in the upper half and lower half planes, respectively. They are further bounded at infinity, and coincide on the real axis $z=f$, where there is a single pole of order one located at $z=0$.

An application of the analytic continuation theorem~\cite{RudinBible} will allow {\em to glue together} the two functions in the upper and lower half planes, obtaining a function which is analytic in the whole plane, except for the single pole of order one at the origin. The generalized Liouville theorem~\cite{RudinBible} defines the only admissible form that such a function can assume: $c/z$, where $c$ is a constant to be determined\footnote{Actually, according to the generalized Liouville theorem, the overall function should be equal to $c_0+c_1/z$. On the other hand, we are looking for a solution $U(f)$ in the class of the functions which vanish at infinity, implying $c_0=0$.}.

Restricting to the real-axis only, we finally get:
\beqa
&&\frac{V^{+}(f)}{1-K(f)} + {[W(f) e^{-i \pi\delta f}]}^{+}=\frac{c}{f}\nonumber\\
&&\frac{V^{-}(f)e^{-i 2\pi\delta f}}{1-\bar{K}(f)} + {[W(f) e^{-i \pi\delta f}]}^{-}=\frac{c}{f}.
\label{eq:finaleq1}
\eeqa
Computing $c$ is made possible by the condition that the sought $u(t)$ should be a probability density function, which is equivalent to $U(0)=1$, or $V^+(0)-V^-(0)=1-\Omega(0)$.
Using eqs.~(\ref{eq:finaleq1}), we can write
\beqa
V^{+}(f)&=&[1-K(f)]{[W(f) e^{-i \pi\delta f}]}^{+}+c \frac{1-K(f)}{f}\nonumber\\
\frac{V^{-}(f)}{e^{i2\pi\delta f}}&=&[1-\bar{K}(f)]{[W(f) e^{-i \pi\delta f}]}^{-}+c \frac{1-\bar{K}(f)}{f},\nonumber
\eeqa
that, evaluated at $f=0$, yield
$V^{+}(0)-V^{-}(0)=-i\,4\pi\,c$,
where we used $\lim_{f\rightarrow 0}\frac{1-K(f)}{f}=-K^\prime(0)=-i 2\pi$. The condition $V^{+}(0)-V^{-}(0)=1-\Omega(0)$ will thus give
\beq
c=i\,\frac{1-\Omega(0)}{4\pi}.
\label{eq:const}
\eeq

If we repeat the above development by multiplying eq.~(\ref{eq:primaeq}) by the complex exponential $e^{i\pi\delta f}$, we  get a similar result, namely\footnote{The structure of the equation is such that the same values of the constant $c=i[1-\Omega(0)]/(4\pi)$ is obtained.}
\beqa
&&\frac{V^{+}(f)e^{i 2\pi\delta f}}{1-K(f)} + {[W(f) e^{i \pi\delta f}]}^{+}=\frac{c}{f}\nonumber\\
&&\frac{V^{-}(f)}{1-\bar{K}(f)} + {[W(f) e^{i \pi\delta f}]}^{-}=\frac{c}{f}.
\label{eq:finaleq2}
\eeqa

Putting together eqs.~(\ref{eq:finaleq1}) and~(\ref{eq:finaleq2}), along with the found value of the constant~(\ref{eq:const}), gives the system of equations
\[
\begin{array}{lcll}
\displaystyle{\frac{V^{+}(f)}{1-K(f)} + {[W(f) e^{-i \pi\delta f}]}^{+}}                         &=&i\, \displaystyle{\frac{1-\Omega(0)}{4\pi f}} &(i)\\
\displaystyle{\frac{V^{+}(f)e^{i 2\pi\delta f}}{1-K(f)} + {[W(f) e^{i \pi\delta f}]}^{+}}        &=& i\,
\displaystyle{\frac{1-\Omega(0)}{4\pi f}} &(ii)\\
\displaystyle{\frac{V^{-}(f)}{1-\bar{K}(f)} + {[W(f) e^{i \pi\delta f}]}^{-}}                    &=&i\,
\displaystyle{\frac{1-\Omega(0)}{4\pi f}} &(iii)\\
\displaystyle{\frac{V^{-}(f)e^{-i 2\pi\delta f}}{1-\bar{K}(f)} + {[W(f) e^{-i \pi\delta f}]}^{-}}&=& i\,
\displaystyle{\frac{1-\Omega(0)}{4\pi f}} . &(iv)
\end{array}
\]

Solving for $V^{+}(f)/[1-K(f)]$ in equations $(i)$ and $(ii)$ gives
\beqa
\left[W(f) e^{i \pi\delta f}\right]^{+}e^{-i \pi\delta f}-
\left[W(f) e^{-i \pi\delta f}\right]^{+}e^{i \pi\delta f}
\nonumber\\
=
\delta\sinc(\delta f)\,\frac{1-\Omega(0)}{2}.
\nonumber
\eeqa
(Using $(iii)$ and $(iv)$ gives identical results.)

The LHS of the above is equivalent to low-pass filtering of $W(f)$, namely $\int W(\nu)\delta\,\sinc[\delta(f-\nu)]d\nu$, so that, by further using the explicit form~(\ref{eq:Xdef}) of $W(f)$, and the properties of $\Omega(f)$, we get the desired claim.~\hfill$\bullet$

\vspace*{10pt} \noindent

For later use, note that at LHS and RHS of eq.~(\ref{eq:finalfinal0}) appear Fourier transforms of functions that vanish outside the range $[-\delta/2,\delta/2]$. In view of the sampling theorem~\cite{PapoulisSA}, the samples taken at $h/\delta$, $h$ integer, define the whole functions. These samples are
\beqa
\Omega(h/\delta)+
2\,\int{\Omega(\nu)\Re\left\{\frac{K(\nu)}{1-K(\nu)} \right \}
\delta\sinc(\delta\nu-h)d\nu}
\nonumber \\
=\delta\,\frac{1-\Omega(0)}{2}I_h,
\nonumber
\eeqa
where $I_h=1$ for $h=0$ and $I_h=0$ otherwise.

Also the unknown function $\Omega(f)$ is bandlimited, so that it can be represented by the Shannon series $\Omega(f)=\sum_k \Omega(k/\delta)\sinc(\delta f-k)$.
Substituting into the above equation we get
\beq
\sum_k A_{hk} \Omega(k/\delta)=\frac\delta 2 I_h, \label{eq:series}
\eeq
where
\beq
\hspace*{-7pt}
\begin{array}{l}
A_{00}=1+\displaystyle{\frac \delta 2}+2\,\int \Re\left\{\frac{K(\nu)}{1-K(\nu)}\right\}\delta\sinc^2(\delta\nu)d\nu \\
A_{kk}=1+2\,\int \Re\left\{\frac{K(\nu)}{1-K(\nu)}\right\}\delta\sinc^2(\delta\nu-k)d\nu, \; k\neq 0 \\
A_{hk}=2\,\int \Re\left\{\frac{K(\nu)}{1-K(\nu)}\right\}\delta\sinc(\delta\nu-h)\sinc(\delta\nu-k)d\nu,\,  h\neq k
\end{array}
\label{eq:AjkFour}
\eeq
having used the orthogonality property of the sinc functions.

Thanks to the results of appendix~\ref{app:Linear}, the above integrals in the Fourier domain can be expressed in the time domain as given in  eqs.~(\ref{eq:A00})--(\ref{eq:symm}).
We are now in the position of proving the remaining claims.

	\subsection{Proof of Theorem 3}

Let us consider only the term $h=0$ in~(\ref{eq:series}):
\beq
A_{00}\Omega(0)+\sum_{k\neq 0} A_{0k}\Omega(k/\delta)=\frac{\delta}{2}.
\label{eq:mequalzero}
\eeq
The rationale behind the approximation of Theorem~3 amounts to neglect the cross-terms $A_{0k}$, for $k\neq 0$, which can be easy understood by considering the two limiting regimes.

Consider first $\delta \ll 1$. By triangle inequality
\[
|A_{0k}|\leq\, \frac{2}{\delta}\int_0^\delta m(t)  \, dt \approx 0
\]
where the approximation follows by $m(0)=0$.

As to $\delta \gg 1$, from a renewal theorem for interarrivals with finite second moment~\cite{ross-book}, we know that
\beq
\lim_{t\rightarrow\infty} [m(t)-t]=\frac{\gamma-1}{2}.
\label{eq:renth}
\eeq
Thus, from~(\ref{eq:symm0}) we write, for $k\neq 0$
\[
A_{0k}=(-1)^{k}\, \frac{2}{\delta}\int_0^\delta \left[m(t)-t-\frac{\gamma-1}{2}\right] \,\cos(2\pi k t/\delta) \, dt,
\]
which follows by $\int_0^\delta t\,\cos(2\pi k t/\delta)=0$. Then, by triangle inequality
\[
|A_{0k}|\leq\, \frac{2}{\delta}\int_0^\delta \left|m(t)-t-\frac{\gamma-1}{2}\right|  \, dt \approx 0,
\]
where the last approximation is a consequence of the Ces\'aro mean theorem and eq.~(\ref{eq:renth}).

We note also that $|\Omega(k/\delta)|<\Omega(0)$ for all $k\neq 0$, and consistently we neglect all terms with $k\neq 0$ in eq.~(\ref{eq:mequalzero}). Solving for $\Omega(0)$ is now possible:
\[
\Omega(0)\approx\frac{\delta}{2 A_{00}}
=\frac
{\delta}
{1-\displaystyle{\frac \delta 2 +\displaystyle{\frac{2}{\delta}\int_0^\delta m(t)dt}}},
\]
yielding, in view of eq.~(\ref{eq:inthethe}), the desired result.

\subsection{Proof of  Corollary 1}
We know that the true embedding capacity tends to unity as $\delta$ diverges. In order to quantify the convergence rate, we consider the limiting behavior of
\beq
1-C=
\frac{1+\frac{2}{\delta}\int_0^\delta m(t)\,dt-\delta}
{1+\frac{2}{\delta}\int_0^\delta m(t)\,dt}.
\label{eq:oneminus}
\eeq
Now,
\[
\frac{2}{\delta}\int_0^\delta m(t)\,dt=
\frac{2}{\delta}\int_0^\delta [m(t)-t]\,dt + \delta
\sim \gamma-1+\delta,
\]
by simple application of the Ces\'aro mean theorem and of the renewal theorem used before, see~(\ref{eq:renth}).
From~(\ref{eq:oneminus}), we get the desired result:
\[
\lim_{\delta \rightarrow \infty} [1-C] \delta= \gamma .
\]

\subsection{Proof of Theorem 4}
Let us consider the series in eq.~(\ref{eq:series}). By truncating it to $2 N +1$ terms, we get the following representation, with $h,k\in[-N,N]$,
\[
\sum_{k=-N}^N A_{hk} \Omega(k/\delta)=\frac \delta 2\,I_h.
\]
Let $\bA$ denote the matrix made of the $A_{hk}$'s.
Recalling that we are not interested in computing the whole function $\Omega(f)$, but just its value at the origin, we get
\[
\Omega(0)=\{\bA^{-1}\}_{00} \, \frac{\delta}{2}.
\]
It is possible to explicit the solution for $N=1$. Using the symmetries involved, we easily get
\[
\{\bA^{-1}\}_{00}=\frac{1}{A_{00}+2\,\frac{A_{01}^2}{A_{01}-A_{11}}},
\]
which, along with eq.~(\ref{eq:inthethe}), yields the desired result~(\ref{eq:lineartheorem}).

\section{Examples}
\label{sec:appl}

The analytical expression of $C$ provided by Theorem~3 turns out to be quite accurate for virtually all the interarrival distributions used in our simulation studies, many of which are typical of network applications. Part of these extensive computer investigations are summarized in Sects.~\ref{sec:closedform} and~\ref{sec:other}.
In fact, to find an example where the refinements offered by Theorem~4 provide meaningful improvements over $C$, we need to choose carefully the kind of interarrival distributions, as discussed later.

\subsection{Examples with Capacity in Closed Form} \label{sec:closedform}
We start with studying some well-known interarrival distributions, for which the renewal function is available in closed form.

\begin{itemize}
\item
{\em Erlang family}
\\
The Gamma random variable u-PDF is
\beq
k(t) = \xi  \frac{(\xi  t)^{\xi-1}}{\Gamma(\xi)} e^{-\xi  t}, \qquad t \ge 0,\qquad \xi>0.
\label{eq:gammapdf}
\eeq
When the parameter $\xi$ is integer, the interarrival distribution belongs to the Erlang family, a case for which the renewal function has been computed in closed form~\cite{MTR}
\[
m(t)=t +  \sum_{h=1}^{\xi-1}  \, \frac{\theta^h}{\xi \, (1-\theta^h)} \,
\left (1- e^{-\xi (1-\theta^h)t } \right ),
\]
with $\theta = e^{i \frac{2 \pi}{\xi}}$.

The (approximation of the) embedding capacity is accordingly
\[
C= \frac{\lambda\Delta}{1 + \lambda\Delta + \displaystyle {2 \sum_{h=1}^{\xi-1} \frac{\theta^h}{\xi(1-\theta^h)} \left ( 1- \frac{1-e^{-\xi(1-\theta^h)\lambda\Delta}}{\lambda\Delta \, \xi(1-\theta^h)}\right ) }}.
\]
\item
{\em Weibull distribution}
\\
The u-PDF for the Weibull random variable is
\beq
k(t) =
\left(\frac{b}{\sigma}\right)
\left(\frac{t}{\sigma}\right)^{b-1}
\,e^{-(t/\sigma)^b}, \qquad t \ge 0,\qquad b>0
\label{eq:weibpdf}
\eeq
where $\sigma=[\Gamma(1+1/b)]^{-1}$.
The pertinent u-RF is~\cite{WeibRenFun}:
\[
m(t)=\sum_{n=1}^\infty
\displaystyle
{
\frac
{(-1)^{n-1}\,a_n\,\left[\Gamma(1+1/b)\,t\right]^{n\,b}}
{\Gamma(1+n\,b)}
},
\]
where the coefficients $a_n$ are defined recursively by
\[
a_1=\alpha_1\dots
a_n=\alpha_n-\sum_{j=1}^{n-1}\alpha_j\,a_{n-j},
\]
with
\[
\alpha_n=\frac{\Gamma(1+n\,b)}{n!}.
\]
This yields
\beq
C= \frac{\lambda\Delta}{1+2\,
\displaystyle{
\sum_{n=1}^\infty
\displaystyle
{
\frac
{(-1)^{n-1}\,a_n\,\left[\Gamma(1+1/b)\,\lambda\Delta\right]^{n\,b}}
{\Gamma(1+n\,b)\,(n\,b+1)}
}
}
}.
\label{eq:WeibCap}
\eeq
\item
{\em Uniform distribution}
\\
The u-RF for uniform random variables can be obtained by iteratively solving the renewal equation~\cite{ross-book}, yielding
\[
m(t)= \left \{
\begin{array}{ll}
e^{ t/2} -1 & 0 \le t \le 2 \\
e^{ t/2} -1 - \left (\frac{ t}{2}-1 \right ) \, e^{ t /2-1}  & 2  \le t \le 4  \\
\dots & \dots
\end{array}
\right .
\]
with similar expressions for successive intervals of length $2$.
The resulting capacity is
\[
C=
\left \{
\begin{array}{ll}
\frac{( \lambda\Delta)^2}{4 e^{\frac{ \lambda\Delta}{2}} -  \lambda\Delta -4 }
\,,
&\lambda \Delta \in[0,2] \\
\\
\frac{ (\lambda\Delta)^2}{4 e^{ \frac{\lambda\Delta}{2}} + 2 e^{\frac{\lambda\Delta}{2}-1} (4- \lambda\Delta) -  \lambda\Delta -8 }
\,,
&\lambda \Delta \in[2,4] \\
\\
\dots & \dots
\end{array}
\right .
\]
\end{itemize}

\begin{figure}
\centerline{\includegraphics[height=.28\textheight]{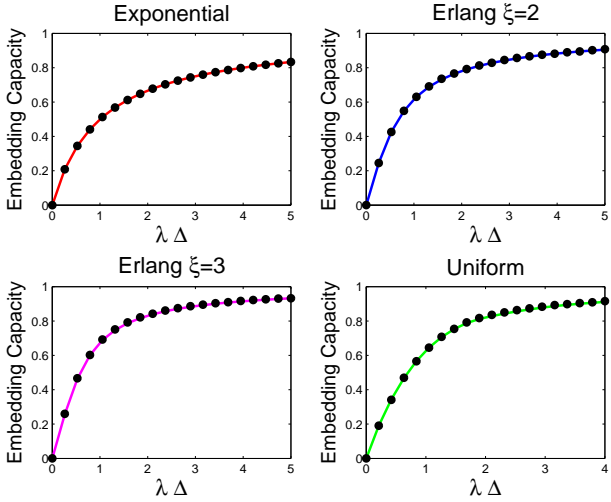}}
\caption{Examples of traffic models for which the renewal function admits simple closed form. Dots refer to computer simulations of the embedding capacity and lines refer to analytical formulas.}
\label{fig:ClosedForm}
\end{figure}

In Fig.~\ref{fig:ClosedForm} the above expressions for the capacity are compared to numerical simulations. We see that the matching between the approximate analytical formulas and the results of computer simulations is excellent.

\subsection{Other Distributions} \label{sec:other}

Even when the renewal function is not known explicitly, there exist many numerical ways to compute that. Some methods exploit the definition of the renewal function in terms of interarrival distribution~\cite{ross-book}, other approaches are based on the interarrival density, and even others exploit the Fourier domain.

\begin{figure}
\centerline{\includegraphics[height=.27\textheight]{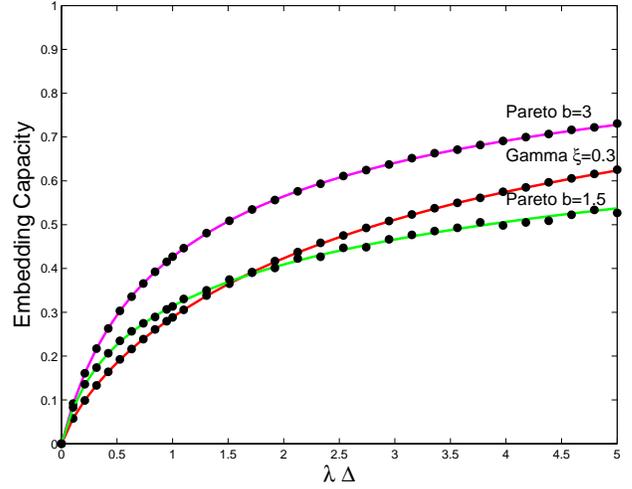}}
\caption{Examples of different traffic models. Continuous curves refer to the approximation $C$ in Theorem~3, eq.~(\ref{eq:finaltheorem}), while dots are obtained by computer simulations.}
\label{fig:OtherDist}
\end{figure}

For instance, let us consider again the Gamma family~(\ref{eq:gammapdf}), and assume that $\xi=0.3$.
In this case, it is particularly convenient to use the Fourier domain expression for $A_{00}$ given in the first equation of~(\ref{eq:AjkFour}). Computing numerically the involved integral, we get the capacity plotted in Fig.~\ref{fig:OtherDist}.

A case of special interest for network applications due to its tail behavior is the Pareto interarrival distribution, whose u-PDF is
\beq
k(t)=\frac{b/(b-1)}{(1+\frac{t}{b-1})^{b+1}}.
\label{eq:paretopdf}
\eeq
Figure~\ref{fig:OtherDist} shows the embedding capacity, again obtained by numerical integration of $A_{00}$ in~(\ref{eq:AjkFour}), for the Pareto distribution
This distribution exhibits finite second moment whenever the shape parameter $b>2$.
Accordingly, in the numerical simulation, we first test the case $b=3$, which falls in the class considered in the assumptions of our theorems, see Fig.~\ref{fig:OtherDist}. Then, we explore by simulation a case with infinite second moment, that is, $b=1.5$, and Fig.~\ref{fig:OtherDist} reveal that the accuracy of the formula is still excellent.

In all the cases examined so far, there is no doubt that the expression $C$ is quite accurate for any practical purposes.
We would like to present an example in which the analytical formula~(\ref{eq:finaltheorem}) of Theorem~3 is less accurate and the following shifted exponential distribution offers this opportunity.
Let us consider the following u-PDF for the interarrivals:
$k(t)=\frac{1}{1-a}\,e^{-\frac{t-a}{1-a}}$, for $t\geq a$, and $0<a<1$.

\begin{figure}
\centerline{\includegraphics[height=.27\textheight]{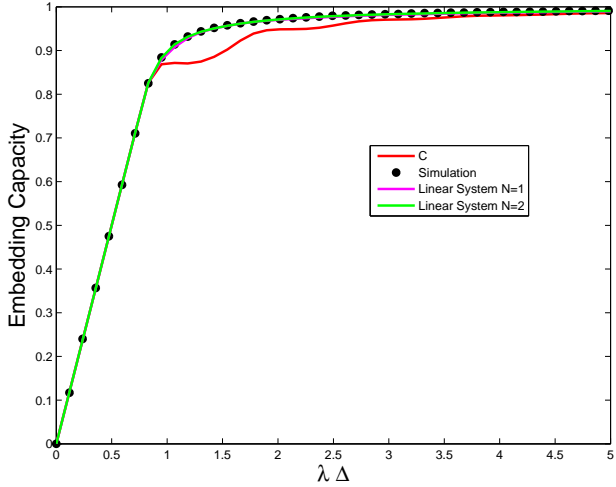}}
\caption{Example of a shifted exponential distribution, with $a=0.8$. Dots are obtained by computer simulations, while continuous curves refer to the different analytical approximations for $C^\ast$ in Theorems~3 and~4. Specifically, we display $(i)$ $C$, $(ii)$ the linear system solution with $N=1$ of eq.~(\ref{eq:lineartheorem}), and $(iii)$ the linear system solution for $N=2$. The latter two curves are superimposed.}
\label{fig:ShiftedExpo}
\end{figure}

The embedding capacity is displayed, together with the simulated data, in Fig.~\ref{fig:ShiftedExpo}.
As it can be seen, the agreement is perfect in the range $\lambda\Delta<a$ and is quite good for large $\lambda\Delta$; this is expected in view of Remark 2, and the arguments used in the proof of Theorem~3. However, for intermediate values of the product $\lambda\Delta$, the approximation $C$ is not satisfying.

Thus, we resort to the refined approximations offered by Theorem~4, and the results are shown again in Fig.~\ref{fig:ShiftedExpo}, where the solutions obtained by using $N=1$ (that is, eq.~(\ref{eq:lineartheorem})), and $N=2$  (this case being solved numerically), are displayed. As it can be seen, the partial inaccuracy of the approximation $C$ is remediated with the adoption of eq.~(\ref{eq:lineartheorem}). Adding more terms (i.e., $N>1$) gives negligible improvements.

\section{Ordering of Embedding Capacities}
\label{sec:ordering}
In this section we show how the embedding capacity $C$ can be used for comparing different renewal processes in terms of their embedding capabilities.
Let $X_1$ and $X_2$ be two non-negative random variables with the same average value $\E[X_1]=\E[X_2]=1/\lambda$, and with cumulative distribution functions denoted by $F_1(\cdot)$ and $F_2(\cdot)$, respectively.
The following definitions and results are classical in stochastic order literature, and can be found in~\cite{stochastic-order,ross-book}.

\vspace*{5pt} \noindent
\textsc{Definition 7}
{\em
(Variability or convex ordering) The random variable $X_1$ is less variable than $X_2$, written $X_1 \le_{v} X_2$, if
\beq
\E[\phi(X_1)]
\leq
\E[\phi(X_2)]
\textnormal{ for all convex functions } \phi:{{\mbbR}\rightarrow{\mbbR}},
\label{eq:cvx}
\eeq
provided that the expectations exist.}~\hfill$\diamond$

\vspace*{3pt} \noindent
\textsc{Known results}~\cite{stochastic-order}
{\em (Sufficient and Necessary Conditions for convex ordering)
For non-negative random variables $X_1$ and $X_2$, with $\E[X_1]=\E[X_2]=1/\lambda$, the condition $X_1 \le_{v} X_2$ is equivalent to each of the following:
\beqa
\qquad \int_0^x \overline{F}_1(t)dt
&\geq&
\int_0^x \overline{F}_2(t)dt,\textnormal{ for all }x,
\label{eq:altord}\\
\qquad \quad L_1(p)&\geq& L_2(p),\textnormal{ for all }p\in[0,1].
\label{eq:Lorenz}
\eeqa
In the above,  $L_1(p)$ and $L_2(p)$ are the so-called Lorenz curves of the random variables $X_1$ and $X_2$ defined as
\[
L_{1,2}(p)=\lambda \int_0^p F^{-1}_{1,2}(u)du,\textnormal{ for all }p\in[0,1].
\]
}

Intuitively, $X_1 \le_{v} X_2$ if $X_1$ gives less weight to the extreme values with respect to $X_2$. One way to get this is just to ensure that $\E[\phi(X_1)] \leq \E[\phi(X_2)]$ for convex $\phi$, as stated in~(\ref{eq:cvx}).
That's why this kind of stochastic ordering is also known as convex ordering. It is also obvious that
$X_1 \le_{v} X_2 \Rightarrow \VAR[X_1] \le \VAR[X_2]$, and hence $X_1$ has a dispersion index smaller than or equal to that of $X_2$, a fact that plays a major role in the regime of $\Delta\gg 1/\lambda$, as seen in Corollary~1.

The following theorem formally relates the classical concept of variability ordering to the embedding capacity in a straightforward and intuitive way:  {\em less variable interarrivals yield a larger embedding capacity}.

\vspace*{5pt}
\noindent
\emph{\textsc{Theorem 5} Let $C_1$ and $C_2$ be the approximate embedding capacities for i.i.d. renewal processes with interarrival distribution $X_1$ and $X_2$, respectively.
Then
\[
X_1{\le}_{v} X_2 \Rightarrow C_1 \ge C_2.
\]
}~\hfill$\diamondsuit$
\vspace*{3pt}

\noindent \emph{Proof.}
The u-RF's of $X_1$ and $X_2$ can be represented as~\cite{ross-book}
\beq
m_1(t)=\sum_{i=1}^\infty \Pr \left\{\lambda S^{(1)}_i \le t\right\},\quad
m_2(t)=\sum_{i=1}^\infty \Pr \left\{\lambda S^{(2)}_i \le t\right\}.
\label{eq:siamoserie}
\eeq
Let us focus on the single terms of the series. By assumption $X_1 {\le}_{v} X_2$, implying, in view of~(\ref{eq:altord}),
\[
\int_0^{\lambda\Delta} \Pr \left\{\lambda S^{(1)}_1 \le t\right\} \, dt \leq \int_0^{\lambda\Delta} \Pr \left\{\lambda S^{(2)}_1 \le t\right\} \, dt.
\]
Since the variability ordering is closed under convolution (see e.g.,~\cite{stochastic-order})
\[
\int_0^{\lambda\Delta} \Pr \left\{\lambda S^{(1)}_n \le t\right\} \, dt \leq \int_0^{\lambda\Delta} \Pr \left\{\lambda S^{(2)}_n \le t\right\} \, dt.
\]
This implies
\[
\int_0^{\lambda\Delta} \sum_{i=1}^n \Pr \left\{\lambda S^{(1)}_i \le t\right\} \, dt
\le
\int_0^{\lambda\Delta} \sum_{i=1}^n \Pr \left\{\lambda S^{(2)}_i \le t\right\} \, dt.
\]
Applying Beppo Levi's monotone convergence theorem~\cite{billingsley-book2}, we are legitimate to exchange integration and limit, yielding
\[
\int_0^{\lambda\Delta} m_1(t) \, dt
\le
\int_0^{\lambda\Delta} m_2(t) \, dt
\]
which, in the light of eq.~(\ref{eq:siamoserie}), gives $C_1\ge C_2$.~\hfill$\bullet$

\subsection{Ordering w.r.t. Poisson}
It is of special interest to compare the given renewal process to Poisson traffic, and this can be conveniently done by means of our analytical approximation.
To do so, let us define two special categories of interarrival distributions.

\vspace*{5pt} \noindent
\textsc{Definition 8}
{\em (NBUE/NWUE classes) A non-negative random variable $X$ is called New Better than Used in Expectation (NBUE) or New Worse than Used in Expectation (NWUE) if~\cite{ross-book}:
\beqa
&\textnormal{NBUE} \quad&\E[X-s|X>s]\leq\E[X]\qquad \forall s\geq 0,\nonumber\\
&\textnormal{NWUE} \quad&\E[X-s|X>s]\geq\E[X]\qquad \forall s\geq 0. \nonumber
\eeqa
}
Due to the absence of memory, the exponential distribution is such that $\E[X-s|X>s]=\E[X]$, and it belongs to both classes.

\vspace*{5pt} \noindent
\emph{\textsc{Corollary 2} (Capacity Ordering in NBUE/NWUE classes)
Let $C_{\mbox{\small NBUE}}$, $C_{\mbox{\small NWUE}}$, and $C_{\mbox{\small exp}}$ denote the embedding capacities given by~(\ref{eq:finaltheorem}) for interarrivals from the NBUE class, the NWUE class, and the exponential distribution. The following relationship holds:
\[
C_{\mbox{\small NWUE}}\leq C_{\mbox{\small exp}}\leq C_{\mbox{\small NBUE}}.
\]
}~\hfill$\diamondsuit$
\vspace*{3pt}

\noindent \emph{Proof.}
Thanks to Proposition 9.6.1 in~\cite{ross-book}, the NBUE (resp.\ NWUE) distributions can be shown to be less (resp.\ more) variable than the exponential, implying the claimed result as a direct consequence of Theorem~5.~\hfill$\bullet$

\begin{table}[h]
\begin{center}		
\begin{tabular}{||c|c|c||}
\hline
 & u-PDF $k(t)$  & Ordering Relationship\\
\hline
GAMMA&  Eq.~(\ref{eq:gammapdf})  &$\xi_1 \geq \xi_2\Rightarrow C_1\geq C_2.$\\
\hline
WEIBULL&  Eq.~(\ref{eq:weibpdf})  &$b_1 \geq b_2\Rightarrow C_1\geq C_2.$\\
\hline
PARETO&  Eq.~(\ref{eq:paretopdf})  &$b_1 \geq b_2\Rightarrow C_1\geq C_2.$\\
\hline
LOGNORMAL&  Eq.~(\ref{eq:lognpdf})  &$\sigma_1 \leq \sigma_2\Rightarrow C_1\geq C_2.$\\
\hline
\end{tabular}
\end{center}
\caption{Summary of the relationships between classical and embedding capacity ordering for typical distributions.}
\label{tab:tavola}
\end{table}

\subsection{Ordering within the same distribution class}
The relationship between embedding capacity ordering and classical ordering of random variables allows easy comparison of distributions within the same class.

For Gamma and Weibull random variables, it has been shown that the Lorenz curves are monotonically increasing with the shape parameters~\cite{Wilfling}.
Thus, larger shape parameters give higher embedding capacities.
It is also easy to evaluate the Lorenz curve of the Pareto random variable with a u-PDF given in~(\ref{eq:paretopdf})
\[
L(p)=b\,(1-p)\,[1-(1-p)^{1-1/b}] + p,
\]
as well as that of the Lognormal random variable
\[
L(p)=\Phi(\Phi^{-1}(p)-\sigma),
\]
whose u-PDF is
\beq
k(t)=\frac{1}{\sqrt{2\pi\sigma^2}\,t}\exp\left\{-\frac{(\log t +\sigma^2/2)^2}{2\sigma^2}\right\}, \qquad t \ge 0.
\label{eq:lognpdf}
\eeq
Both functions exhibit monotonic behavior with respect to $b$ and $\sigma$, respectively.

Therefore, using eq.~(\ref{eq:Lorenz}) allows easy (convex) ordering of the interarrivals, which in turns induces an ordering of the embedding capacities thanks to Theorem 5. The results are summarized in Table~\ref{tab:tavola}.

\begin{figure}
\centerline{\includegraphics[height=.27\textheight]{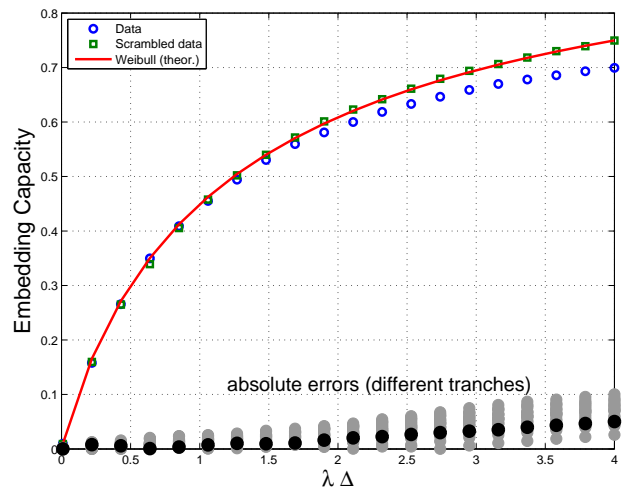}}
\caption{Embedding capacity curve of Telnet data, for a pair of tranches selected as described in the main text. 
In the lower part of the plot, the absolute error between empirical and theoretical capacity is displayed, for a broader set of different tranches.}
\label{fig:TelnetCap}
\end{figure}

\section{Experiments with Real Network Traces}
\label{sec:realtraces}
In this section, we present some numerical tests  run on real traces. Specifically, we downloaded the TCP packet arriving times (traces {\em lbl-tcp-3.tcp} and {\em lbl-pkt-4.tcp}) gathered at the Lawrence Berkeley Laboratory, that were originally used in~\cite{PaxsonFloyd}.
Following~\cite{PaxsonFloyd} and~\cite{He-Tong-IT}, we extract packets corresponding to Telnet connections (obtained from hearing communication on port 23). The pipeline for the real data processing is as follows.
\begin{itemize}
\item
The inspected traces correspond to traffic patterns collected in two different days.
We consider the {\em aggregate} traffic, that is, we do not extract informations pertaining to the single hosts.
\item
We emulate the scenario of two mutually independent point processes, by using two tranches of $10^4$ packets each, extracted from traces {\em lbl-tcp-3.tcp} (source node) and  {\em lbl-pkt-4.tcp} (relay node).
\item
By means of a moving average filter over $10^4$ packets, we select over the two traces candidate tranches having comparable rates\footnote{With this selection procedure,  the tranches extracted  from a given trace might also overlap. 
Obviously, this does not alter our analysis, in that we only need independence between the source tranche (extracted from trace {\em lbl-tcp-3.tcp}), and the relay tranche (extracted from the {\em independent} trace {\em lbl-pkt-4.tcp}). }. Without loss of generality, we scale the data by dividing the interarrivals by the sample mean computed over the union of the two tranches.
\item
We run the BGM algorithm on the selected tranches, with fixed (dimensionless) observation time $t=9000$.
\item
We also run the BGM algorithm {\em after scrambling} the interarrivals, in order to remove statistical dependencies between them, namely,  to enforce the renewal assumption. This is purely for testing the accuracy of the found formulas. 
\end{itemize}
In order to compute theoretical capacities, we need a candidate distribution for the interarrivals. 
We accordingly fit the empirical interarrival CDF of each tranche, and find that the Weibull distribution works generally  well, that is perhaps not unexpected, see, e.g.,~\cite{Norros} and~\cite{Papagiannakietal}. 

Consider now the capacity curves in Fig.~\ref{fig:TelnetCap}.
The experimental curves for capacity refer to one pair of tranches where the Weibull fit is accurate, and the two empirical CDFs are close to each other, complying with the assumption of  identical distribution across nodes.
The theoretical curve is drawn by~(\ref{eq:WeibCap}), where the shape parameter $b$ is computed over the union of the two tranches. 

For the scrambled data the theoretical approximation $C$ is excellent.
As to the (non-scrambled) real data, a first evidence is that, up to values of $\lambda\Delta$ in the order of unit, the curve matches the theoretical approximation well. 
On the other hand, a discrepancy emerges at larger values of the product $\lambda\Delta$, due to possible dependencies among the interarrivals. 

A more complete picture is obtained by applying the above procedure to different tranches, irrespectively of the goodness of the Weibull fit, and of the similarity between the empirical distributions at the two nodes. 
The results of this latter analysis are summarized in the bottom part of Fig.~\ref{fig:TelnetCap}, where the absolute error between the theoretical formula and the empirical capacity is displayed, only for the case of real data. (Again, scrambling reduces the error, this is not shown in the plot.)
The points marked with darker circles refer to the tranche pair used for drawing the capacities displayed in the uppermost part of the plot. 
As it can be seen, the theoretical approximation follows the empirical capacity closely at small $\lambda\Delta$.
Also in this case, a discrepancy is observed for moderately large values of $\lambda\Delta$, with an absolute error staying in the order of $10^{-1}$. 

Summarizing, a main behavior seems to emerge --- that the theoretical predictions are very accurate for real data well modeled by renewal processes, corroborating the whole theoretical machinery for embedding capacity computation, and that the possible statistical dependence among packet interarrivals can be neglected for {\em tight} delay constraints, up to delay values in the order of the mean interarrival time.

\section{Conclusion}
\label{sec:concl}
We consider the problem of matching two independent and identically distributed renewal processes, according to a bounded delay criterion, with applications to communication network scenarios.
We introduce the concept of {\em embedding capacity}, and provide fully analytical tools and approximations to evaluate it, relying upon the Riemann-Hilbert theory. An exact evaluation of the capacity is reduced to a manageable integral equation, that can be solved to any degree of approximation by inverting a highly structured linear system. The main finding, however, is a simple approximated formula of the embedding capacity that involves the renewal function of the underlying processes. The approximation is excellent for virtually all the cases of practical interest that we have investigated, part of which are reported in the paper. Even when this is not strictly true, we provide closed-form solutions for first-order correction.

The  analytical formula highlights the role played by different renewal parameters: 
for large $\lambda\Delta$ only the dispersion index matters, while embedding capacity ordering is induced by the stochastic variability of the underlying interarrivals.

The experimental analysis carried on real network traces reveals that the accuracy of the analytical expression is good for tight delay constraints, up to $\lambda\Delta$ in the order of unit.
For larger delays, a partial inaccuracy is seen, and we show that this should be ascribed to statistical dependencies unavoidably present in real traffic patterns: the renewal model is failing, rather than the proposed analytical approximation.

The abstract concept of matching between point processes arises in a very large number of contexts, and we feel that our findings can represent a contribution to these fields. To broaden further the horizon of potential applications, refinements and improvements of the approach can be considered. These the case of different renewal processes at the two nodes, the extension to non-renewal point processes, to multi-hop flows, and to the case of multiple input/multiple output relays, see~\cite{Venk-He-Tong-IT,He-Tong-IT}.

\appendices
\section{Proof of Theorem 1}
\label{app:Markov}
We first justify the embedding capacity formula~(\ref{eq:CapExpress}). Assume for now that the frequency for $Z_n$ to fall inside the interval $[0,\: \Delta]$ converges a.s. to a constant $p$.
Then since each $Z_n$ outside $[0,\: \Delta]$ represents a chaff point whereas each $Z_n$ inside the interval represents a pair of flow points, we see that the fraction of flow points embedded by BGM converges a.s., and the limit, i.e., the embedding capacity, is given by $2\,p/(1+p)$.

On the other hand, by Theorem 17.1.7 in~\cite{Meyn&Tweedie-book}, if $\{Z_n\}$ is a {\em positive Harris recurrent} Markov chain, then $i)$ $p=\lim_{n\rightarrow\infty}{\frac 1 n}\sum_{j=0}^{n}I_{[0,\: \Delta]}(Z_j)$ exists a.s., where $I_{[0,\: \Delta]}(z)$ is the indicator function,  and $ii)$ $p$ can be computed from the invariant PDF $h(t)$ by $p=\int_0^\Delta h(t)dt$. By definition, if $h(t)$ is the solution to eq.~(\ref{eq:mainEq0}), it will be invariant under the transition (\ref{eq:Markov}), i.e., it is an invariant measure. The positive Harris property of the chain implies that $h(t)$ is unique and finite, and thus can be normalized into a probability measure.
It remains to prove the property of positive Harris recurrence.

First, we show that the Markov chain $\{Z_n\}$ is $\psi$-irreducible~\cite{Meyn&Tweedie-book} (all the sets mentioned in the sequel are Borel). The assumption that BGM can match one pair almost surely implies that the interval $[0,\Delta]$ is accessible from any state almost surely, say $L(z,[0,\Delta])\equiv 1$ $\forall z$~\cite{Meyn&Tweedie-book}. This rules out the cases where the asymptotic fraction of matched points depends on the initial state (where embedding capacity does not exist) and those where the embedding capacity is trivially zero.

\begin{figure}
\centerline{\includegraphics[height=.08\textheight]{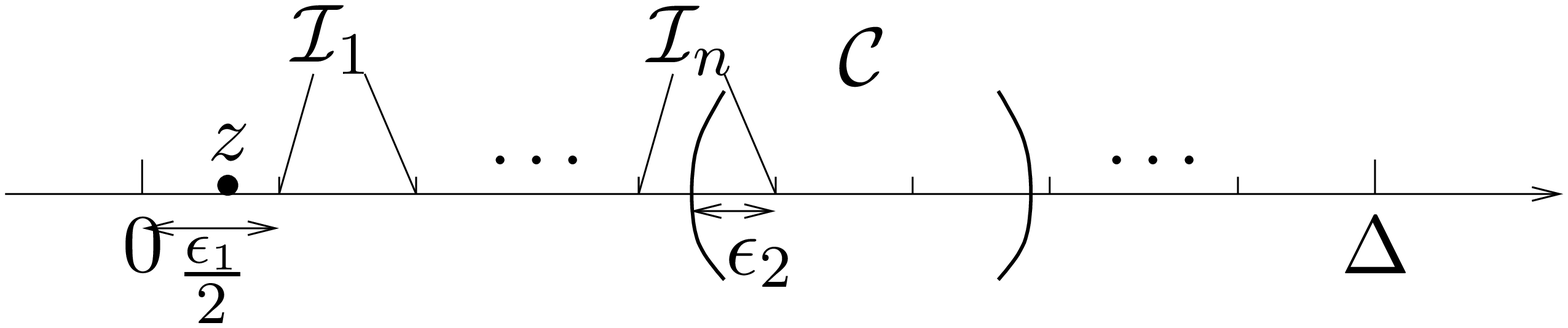}}
\caption{Access ${\cal C}$ from $z$ by hoping through ${\frac{\epsilon_1}{2}}$-segments in $[0,\Delta]$.}
\label{fig:proof1}
\end{figure}

Let $\varphi$ be the Lebesgue measure constrained to $[0,\: \Delta]$, i.e.
$\varphi({\cal A})=\mu({\cal A}\cap [0,\: \Delta])$, where $\mu$ is the Lebesgue measure. Given PDF $f(t)$, there must exist $\epsilon_0>0$ such that $f(t)>\delta_0$ for all $t$ within some interval $[t_0,\: t_0+\epsilon_0]$, and thus
\[f_0(t)=\int_0^{+\infty} f(\tau)f(\tau-t)d\tau > \delta_0^2(\epsilon_0-|t|)\geq \delta_0^2(\epsilon_0-\epsilon_1)\]
for all $t\in [-\epsilon_1,\epsilon_1]$, where $\epsilon_1$ is a constant in $(0,\epsilon_0)$. Let $\delta_1:= \delta_0^2(\epsilon_0-\epsilon_1)$. Partition $[0,\Delta]$ into $m:= \lceil 2\Delta/\epsilon_1\rceil$ segments of length $\epsilon_1/2$, as illustrated in Fig.~\ref{fig:proof1}, such that the transition density from any $z\in[0,\Delta]$ to any point in an adjacent segment is greater than $\delta_1$. For any set ${\cal C}$ with $\varphi({\cal C})>0$, let $\epsilon_2$ be the Lebesgue measure of the minimum intersection between ${\cal C}$ and the ${\frac{\epsilon_1}{2}}$-segments. Let $z$ be an arbitrary point in $[0,\Delta]$ that is $n$ segments away from ${\cal C}$ ($n\leq m-1$) and ${\cal I}_i$ ($i=1,\ldots,n$) be the $i$th segment from $z$ to ${\cal C}$, where ${\cal I}_n$ intersects with ${\cal C}$. The $n$-step transition satisfies
\begin{align}
P^n(z,{\cal C}) &> \int_{{\cal I}_1} f_0(x_1-z)dx_1\int_{{\cal I}_2} f_0(x_2-x_1)dx_2 \cdots \nonumber\\
& \int_{{\cal I}_n \cap {\cal C}} f_0(x_n-x_{n-1})dx_n \nonumber \\
&> \left({\delta_1 \frac{\epsilon_1}{2}}\right)^{n-1}\delta_1\epsilon_2 > 0. \label{eq:Pn(z,C)}
\end{align}
This implies $L(z,{\cal C})>0$ for all $z\in[0,\Delta]$. Moreover, since $L(z,[0,\Delta])\equiv 1$ for all $z$, we have $L(z,{\cal C})>0$ for all $z$. That is, any set with positive $\varphi$ measure is accessible from anywhere within the state space with positive probability, implying that the chain is $\varphi$-irreducible and hence $\psi$-irreducible for a maximal irreducibility measure $\psi$, according to~\cite{Meyn&Tweedie-book}.

Second, we show that $\{Z_n\}$ is Harris recurrent. Since it is $\psi$-irreducible and $L(z,[0,\: \Delta])>0$ for all $z$, by Theorem 5.2.2 in~\cite{Meyn&Tweedie-book}, there exist $k\geq 1$, a nontrivial measure $\nu_k$, and a nontrivial set ${\cal C}_1\subseteq [0,\: \Delta]$ such that ${\cal C}_1$ is $\nu_k$-small, and hence $\nu_{\delta_k}$-petite. For sampling distribution $a(i)=1/m$ ($i=1,\ldots,m$), the transition kernel of the sampled chain from any $z\in [0,\Delta]$ satisfies
\begin{align}
K_a(z,{\cal C}_1) \geq { \frac 1 m} P^n(z,{\cal C}_1) > {\frac 1 m}\left({\delta_1 \frac{\epsilon_1}{ 2}}\right)^{n-1}\hspace{-1.5em}\delta_1\epsilon_2, 
\end{align}
where we apply (\ref{eq:Pn(z,C)}) for ${\cal C}={\cal C}_1$. Since $n\leq m-1$, $K_a(z,{\cal C}_1)> {\frac 1 m} \left({\delta_1\epsilon_1/ 2}\right)^{m-2}\delta_1\epsilon_2$, independent of $z$ for $z\in [0,\Delta]$. Therefore, ${\cal C}_1$ is uniformly accessible using $a$ from $[0,\Delta]$. By Proposition 5.5.4 in~\cite{Meyn&Tweedie-book}, we prove that $[0,\: \Delta]$ is  $\nu_{a*\delta_k}$-petite.
The fact that a petite set $[0,\: \Delta]$ satisfies $L(z,[0,\Delta])\equiv 1$ for all $z$ for a $\psi$-irreducible chain implies Harris recurrence in the light of Proposition 9.1.7 in~\cite{Meyn&Tweedie-book}. 

Finally, we show positivity by drift analysis. Define the function
\[
V(z)=2\lambda\left\{
\begin{array}{lll}
z-\Delta,\qquad &\mbox{if }\; z>\Delta,\\
0,\qquad &\mbox{if }\; 0\leq z\leq\Delta,\\
-z,\qquad &\mbox{if }\; z<0,
\end{array}
\right.
\]
where $1/\lambda$ is the mean interarrival time, and
consider the mean drift defined in~\cite{Meyn&Tweedie-book} as 
\[
dV(z)=
\int P(z,dy)V(y)-V(z),
\]
where $P(z,dy)$ is the transition kernel of the chain. 
Define a set ${\cal C}_2= [-z_0,\: \Delta+z_0]$ for $z_0$ sufficiently large such that $\int_0^{z_0}f(t)tdt-\int_{z_0+\Delta}^\infty f(t)tdt\geq 1/(2\lambda)$. 
For any $z>\Delta+z_0$ we have, after some straightforward manipulations, 
\beqa
dV(z)&=&-2\lambda
\left[
-\int_{z}^{\infty} f(t)(t-z) dt\right.\nonumber\\
&+&\left.\int_0^{z-\Delta} f(t) t dt
+ (z-\Delta)\int_{z-\Delta}^\infty f(t) dt
\right]\nonumber\\
&\leq& -2\lambda
\left[
\int_0^{z_0} f(t) t dt
-\int_{z_0+\Delta}^{\infty} f(t) t dt
\right]\leq -1.\nonumber
\eeqa
The same holds for $z<-z_0$. It is easy to see that, inside the set ${\cal C}_2$, $dV(z)$ can be  bounded by a constant, such that we can write
\beq
dV(z)\leq -1 + b\,I_{{\cal C}_2}(z),
\label{eq:drift}
\eeq
with a suitable choice of $b$. 
Since the petite set $[0,\Delta]$ is uniformly accessible
\footnote{This can be easily shown with the same technique used to prove uniform accessibility of ${\cal C}_1$ from $[0,\Delta]$.}
from ${\cal C}_2$, we can conclude that ${\cal C}_2$ is petite, and eq.~(\ref{eq:drift}) coincides with the drift condition ($iv$) of Theorem 13.0.1 in~\cite{Meyn&Tweedie-book}, whence, further observing that aperiodicity holds, we conclude that $\{Z_n\}$ is positive Harris.~\hfill$\bullet$

\section{Linear system coefficients}
\label{app:Linear}
Let us introduce the so-called {\em renewal density} associated to the renewal function $m(t)$, that is $\rho(t)=d m(t)/dt$.
It is convenient to consider a symmetric version thereof, namely $\widetilde \rho(t)=\rho(t)+\rho(-t)$.
It holds true that the Fourier transform of $\widetilde \rho(t) -1$ is given by
$2\,\Re\left\{\frac{K(f)}{1-K(f)}\right\}$, see~\cite{FellerOrey,Carlsson}.

Let us first consider the term $A_{00}$ in eq.~(\ref{eq:AjkFour}).
In view of Parseval's formula~\cite{BracewellBook}:
\beqa
\lefteqn{2\,\int \Re\left\{\frac{K(\nu)}{1-K(\nu)}\right\}\delta\sinc^2(\delta\nu)d\nu}\nonumber\\
&=&\int_{-\delta}^\delta [\widetilde\rho(t)-1](1-|t|/\delta )dt
=2\,\int_{0}^\delta \rho(t)(1-t/\delta)dt-\delta,\nonumber
\eeqa
where we simply notice that the Fourier transform of the triangular window of width $2\delta$ is $\delta\sinc^2(\delta f)$.
Integration by parts then gives
$2\,\int_{0}^\delta \rho(t)(1-t/\delta)dt=\frac{2}{\delta}\int_0^\delta m(t)dt$,
or
\[
A_{00}=1-\frac \delta 2 +\frac{2}{\delta}\int_0^\delta m(t)dt.
\]

As to the evaluation of $A_{kk}$ in eq.~(\ref{eq:AjkFour}), $k\neq 0$, it suffices to use the shift property of the Fourier transform, yielding
\beqa
\lefteqn{2\,\int \Re\left\{\frac{K(\nu)}{1-K(\nu)}\right\}\delta\sinc^2(\delta\nu-k)d\nu}\nonumber\\
&=&\int_{-\delta}^\delta [\widetilde\rho(t)-1](1-|t|/\delta)\,\cos(2\pi k t/\delta)dt\nonumber\\
&=&2\,\int_{0}^\delta \rho(t)(1-t/\delta)\,\cos(2\pi k t/\delta)dt, \nonumber
\eeqa
that integrated by parts gives
\beqa
\lefteqn{A_{kk}=1+ \frac 2 \delta \int_0^\delta m(t) [\cos(2\pi k t/\delta)\,dt} \nonumber \\
&&+\, 2 \pi k \, ( 1-t/\delta)\, \sin (2 \pi k t / \delta)  ] \, dt. \nonumber
\eeqa

Finally focusing on the terms $A_{hk}$ in eq.~(\ref{eq:AjkFour}), $h\neq k$,
it suffices to consider the even part of $\delta\sinc(\delta f-h)\sinc(\delta f-k)$, whose inverse Fourier transform is
\beqa
\lefteqn{
\frac{1}{\delta}\,
\Re
\left\{
\int_{-\delta/2}^{\delta/2}
e^{-i 2\pi(h-k)\frac{\tau}{\delta}}
e^{-i 2\pi k\frac{t}{\delta}}
\Pi\left(\frac{t-\tau}{\delta}\right)
d\tau
\right\}
}
\nonumber\\
&=&
\int_{-1/2}^{1/2}
\cos[\,2\pi (h-k)\tau + 2\pi k t/\delta\,]
\,\Pi(\tau-t/\delta)
\,d\tau,\nonumber
\eeqa
$\Pi(t)$ being the rectangular window of width $1$.
The integral is zero for $|t|>\delta$. For $t\in (0,\delta)$ we have
\beqa
\lefteqn{\int_{t/\delta-1/2}^{1/2}
\cos[2\pi (h-k)\tau + 2\pi k t/\delta ]
d\tau}\nonumber\\
&=&
\frac{(-1)^{h-k}}{2\pi (h-k)}
[\,\sin(2\pi k t/\delta )
-\sin(2\pi h t/\delta )
\,] .\nonumber
\eeqa
This gives
\beqa
\lefteqn{A_{hk}=
\frac{(-1)^{h-k}}{2\pi (h-k)}
\,2\,\int_0^\delta
\rho(t)
[\,\sin(2\pi k t/\delta )
-\sin(2\pi h t/\delta )
\,]\,dt} \nonumber \\
&& \hspace*{-12pt}=\frac{(-1)^{h-k}}{(h-k)}\, \frac{2}{\delta}\int_0^\delta m(t) [h\,\cos(2\pi h t/\delta)
-\, k\,\cos(2\pi k t/\delta )]dt \nonumber ,
\eeqa
where the latter is obtained integrating by parts. Equation~(\ref{eq:symm0}) now follows as a special case, whence eq.~(\ref{eq:symm}) is true.


\end{document}